\documentclass[runningheads]{llncs}
\usepackage[T1]{fontenc}
\usepackage{graphicx}
\usepackage{booktabs}
\usepackage[misc]{ifsym}
\usepackage{hyperref}
\usepackage{multirow}
\usepackage{amsmath}
\usepackage{algorithm}
\usepackage{algorithmic}
\usepackage{balance}
\usepackage{xcolor}
\usepackage{makecell}
\usepackage{dsfont}
\usepackage[acronym,nohypertypes={acronym}]{glossaries}
\newacronym{rl}{RL}{\textit{Reinforcement Learning}}
\newacronym{sla}{SLA}{\textit{Service Level Agreement}}
\newacronym{dbms}{DBMS}{\textit{Database Management Systems}}
\newacronym{dba}{DBA}{\textit{Database Administrators}}
\newacronym{vm}{VM}{\textit{Virtual Machine}}
\newacronym{ml}{ML}{\textit{Machine Learning}}
\newacronym{dnn}{DNN}{\textit{Deep Neural Network}}
\newacronym{tps}{TPS}{\textit{Transactions Processed per Second}}
\newacronym{qps}{QPS}{\textit{Queries Processed per Second}}
\newacronym{llm}{LLM}{\textit{Large Language Model}}
\newacronym{pca}{PCA}{\textit{Principal Component Analysis}}
\newacronym{ddpg}{DDPG}{\textit{Deep Deterministic Policy Gradients}}
\newacronym{drl}{DRL}{\textit{Deep Reinforcement Learning}}
\newacronym{tco}{TCO}{\textit{Total Cost of Ownership}}
\newacronym{ppo}{PPO}{\textit{Proximal Policy Optimization}}
\newcommand{\corr}{(\Letter)}
\usepackage{mwe}

\begin{document}

\title{Reinforcement Learning based DBMS Buffer Pool Auto-Tuning for Optimal Memory Utilization}

\titlerunning{RL-based DBMS Buffer Pool Auto-Tuning }

\author{Yifan Wang\inst{1,2,3}\corr \and
Patrick Royer\inst{1} \and
Rapha\"el F\'eraud\inst{1} \and
David Delande\inst{1}}

\authorrunning{Y. Wang et al.}

\institute{
Orange, France \\
\email{\{yifan.wang, patrick.royer,raphael.feraud,david.delande\}@orange.com}
\and
Inria, France
\and
Université de Lille, France
}

\maketitle              

\begin{abstract}
Administering \gls{dbms} instances requires \gls{dba} to balance performance in terms of \gls{sla} against resource usage, often prompting RAM over-allocation that wastes memory. We introduce \emph{MicroTune}, an online RL-based buffer adjustment system that minimizes unnecessary memory allocation while ensuring SLA compliance. To identify the most effective RL core, we evaluate multiple algorithms under diverse benchmark workloads, training \emph{MicroTune} on extensive traces of both external metrics (latency, throughput) and internal DBMS metrics (status variables and performance statistics). Experimental results demonstrate that \emph{MicroTune} dynamically adapts buffer sizes to workload fluctuations, outperforming baselines by achieving significant memory savings with fewer SLA violations. These findings underscore the promise of reinforcement learning for adaptive resource management in \gls{dbms} environments.

\keywords{Reinforcement Learning, Database}

\end{abstract}

\section{Introduction}

In recent years, the exponential growth in data volumes~\cite{statistaWorldwideDataCreated} has significantly increased the demands on \gls{dbms} resources and intensified the challenges of \gls{dba}.
As complex, and specialized software designed to manage data efficiently, \gls{dbms} requires substantial computational resources (CPU, RAM, etc.) and the administration of experts.
The administration tasks, particularly database tuning, are crucial as they involve resource allocation and configuration adjustments that directly influence performance~\cite{tuning}. 

The fine-tuning of many DBMS systems presents two primary challenges:
    1) It requires a substantial amount of expert time for each DBMS, as it involves complex analyses of workloads, which vary in query types, thread counts, data structures, and volumes.
    2) Workloads are dynamic, often fluctuating in response to human activity throughout the day. Additionally, as the applications associated with the DBMS evolve, the behavior of the DBMS workload changes accordingly.
Reducing the costs associated with memory over-allocation in DBMS systems is challenging. It requires a dynamic, real-time strategy capable of continuously adjusting RAM allocation to accommodate changing workloads. 
In modern deployment scenarios, such as Container-based~\cite{docker} or Kubernetes-managed clusters~\cite{kubernetes}, memory resources can be reclaimed and reallocated to other services or applications when not needed by the DBMS. Notably, the ability to adjust the memory assigned to a Kubernetes pod was recently introduced in version 1.32~\cite{kubernetes_v1_32_penelope_2024}. 
A dynamic fine-tuning of the RAM allocation can not only improve overall resource utilization but also contribute to lowering the \gls{tco} for the database system.

Prior work has largely focused on one-time tuning rather than dynamic online adaptation. Search-based tuners (e.g., {\sc BestConfig}~\cite{bestconfig}) sample knob configurations but incur high overhead; Bayesian optimization tuners (e.g., {\sc Ottertune}~\cite{ottertune}) leverage ML but depend on extensive training data; and \gls{rl}-based tuners (e.g., UDO~\cite{UDOtune}, {\sc CDBTune}~\cite{CDBtune}) explore the configuration space adaptively, yet remain limited to static settings. Rule-based methods such as \textsc{iBTune}~\cite{ibtune} target buffer minimization under miss-ratio constraints, while {\sc Restune}~\cite{Restune} broadens scope to CPU, RAM, and I/O but requires replayed workloads and slow convergence. Other studies~\cite{Storm2006,Lasch2023,Kuiper2024,Otaki2025} focuse on internal RAM redistribution without changing the DBMS’s total allocation.

Cloud-scale systems often mitigate over-allocation through forecasting based provisioning (e.g., Redshift~\cite{Nathan2024}, Intelligent Pooling~\cite{Ravikumar2024}, Oracle Cloud Infrastructure \cite{oci}), supported by proprietary infrastructures and workload traces. While enterprise-grade cloud databases include built-in resource auto-adjustment, our focus is on free open-source DBMSs, where such mechanisms are absent. \emph{MicroTune} addresses this gap by introducing a lightweight RL-based agent for online buffer tuning, motivated both by vendor lock-in risks and eventual cost considerations~\cite{MYWORK}. 

Building on these insights, we propose \emph{MicroTune}, an online tuner that dynamically adjusts RAM allocation in real time while satisfying \gls{sla} requirements.
\emph{MicroTune} leverages multiple \gls{rl} algorithms. the use of a \gls{rl} algorithms allows adjusting RAM to change of the workload (the variation of the query type, number of connections, database size etc.) by introducing a dependence between the observed context, which captures information about DBMS and workload, and the right decision for increasing or decreasing the needed RAM. 
This approach enables \emph{MicroTune} to efficiently tune system parameters in an automatic, continuous manner, 
that optimizes DBMS resource utilization, reduces \gls{tco}, and alleviates users from the burden of fine-tuning.

\textbf{Contributions.} In this paper, we make the following contributions. 
Section~\ref{env_states_etc} formulates buffer-size optimization under SLA constraints as a \gls{rl} problem and introduces an optimal-policy baseline. Section~\ref{System Overview} presents a modular architecture for data collection, training, policy selection, and real-time deployment. Section~\ref{exps} 
evaluates non-learning baselines and state-of-the-art \gls{rl} algorithms, showing that \gls{rl} consistently outperforms alternatives and adapts buffer allocations to workloads, while in section \ref{online} \emph{MicroTune} is successfully tested on a live database in realtime.
Finally, Section~\ref{conclusion} discusses future directions and concludes.

\section{ \gls{rl} based real-time DBMS memory cache auto-tuning }
\label{env_states_etc}

\subsection {Problem statement}
\label {Problem_statement}

Let $\mathcal {W}$ be the set of possible workloads. $w \in \mathcal {W}$ is the workload run on the database.
The latency depends on the workload $w \in \mathcal {W}$, the buffer size $b \in [B_{min},B_{max}]$, $B_{max} \geq B_{min} \geq 0$, and the noise corresponding to the activity of the telecommunication network and the server that hosts the database. 
Let $l \sim \upsilon(w,b),\text{ } l > 0$ be the random variable that corresponds to the latency, and $\upsilon(w,b)$ its distribution.
$l_t$ denotes the observed latency at time $t$.
The objective is to find with high probability the minimum buffer size $b_w^*$ for the workload $w$ for which the SLA constraint is satisfied ($l_t \leq l^*$):

\begin {equation}
\label {eq:OIA}
 b^*_w = \arg \min_{b \in [B_{min},B_{max}]} \{ b \text { such that, } P( l \sim \upsilon(w,b)) \geq l^*) \leq 1-\delta \}.
\end {equation}

The optimal buffer size $b^*_w$ of the database for the workload $w$ is denoted as the Oracle.

\begin{algorithm}
	\caption{Optimization of buffer size}
	\label{alg:OIA}
	\begin{algorithmic}[1]
        \FOR {$t \leftarrow 1,...,T-1$}
		    \STATE A workload $w_t \in \mathcal {W}$ is executed
		    \STATE The agent chooses a buffer size $b_t$
			\STATE The latency $l_{t}\sim \upsilon_l(w_t,b_t)$ is revealed
		\ENDFOR
	\end{algorithmic}
\end{algorithm}

Algorithm \ref {alg:OIA} summarizes the optimization problem of the buffer size of a database.
In the following, we will formalize this problem as a reinforcement learning problem, where an agent interacts with an unknown environment~\cite {rl} by choosing buffer size $b_t$.

 
\subsection{Environment}
The environment refers to the \gls{dbms} and the workloads. Specifically, the buffer size of the DBMS is the parameter being tuned. The DBMS responds to actions (changes in the buffer cache size) by reflecting changes in its states and in its latency for the executed workload.

\subsection{States}
\label{states}
For choosing the buffer size $b_t$ at time $t$, the agent observes the current state of the database $s_t \in \mathcal {S}$, which reflects the effect of the workload $w \in \mathcal{W}$ for the buffer size $b_t$, where $\mathcal {S}$ is the set of possible states.

In the context of MariaDB, metrics are automatically computed by the DBMS, which offer a comprehensive view of its operational state, covering aspects such as concurrency levels, buffer-pool activity, locking behavior, and row-level insertions. Depending on the DBMS version and configuration, more than 400 distinct metrics are maintained in the Information Schema and Performance Schema. \gls{dba} commonly rely on these indicators to monitor system health~\cite{CDBtune}, and they can be easily retrieved by querying the native DBMS information tables.
It is important to note that some of these metrics are cumulative, so we compute the difference between two consecutive measurements to capture meaningful state observations. The selection of these metrics is aimed at accurately representing the DBMS state while maintaining a manageable dimensionality.  
To achieve this balance, we applied a correlation-based filter—drawing on the methodology from Ottertune~\cite{ottertune}—to remove redundant metrics and retain only the 51 most informative features.

\subsection{Actions}
\label{actions}
Rather than directly choose the buffer size given a database state, we choose to reach the optimal buffer size by increments or decrements of 128MB. 
The rationale behind the fixed increments or decrements of 128MB is to avoid drastic changes to the buffer size, which could negatively impact DBMS performance. Indeed, adjusting the buffer size requires the DBMS to reorganize its buffer pool, which can introduce instability, notably in case of significant reduction.
Moreover, in practice the workloads do not change at each time step $t$. The DBMS can use multiple time steps to optimize the buffer size.
The goal is to implement a moderate and gradual adjustment strategy to ensure system stability while optimizing performance. The set $\mathcal {A}$ is defined with three actions: 
\begin{enumerate}
    \item \textit{Down}: Decrease the buffer pool size by 128MB, 
    \item \textit{Stay}: Maintain the current buffer pool size, 
    \item \textit{Up}: Increase the buffer pool size by 128MB. 
    \end{enumerate}

\subsection{Policy} 
A policy $\pi \in \Pi: \mathcal{S}\rightarrow \mathcal {A}$ returns the action to play (\textit{Down}, \textit{Stay}, Up) given the current state of the database, where $\Pi$ is the set of policies that depends on the used RL algorithm. When the reward is received, the policy is updated. Any RL algorithms (probabilistic or deterministic) can be used for updating the policy, which aims to handle the optimization of buffer size (Algorithm \ref {alg:OIA}).

\subsection {Optimal Policy}

To take into account the action space, we define the optimal policy $\pi^* \in \Pi$ that at each step, picks the action in 
$\mathcal A = \{\text{Down},\,\text{Stay},\,\text{Up}\}$,
which moves $b_t$ closest to the oracle’s $b^*_w$ (Algorithm \ref{alg:Oracle}).

\begin{algorithm}
    \caption{Optimal Policy $\pi^*$}
    \label{alg:Oracle}
    \begin{algorithmic}[1]

    \STATE Call the Oracle for obtaining the optimal buffer size $ b^*_{w}$ for workload $w$
        
        \FOR {$t \leftarrow 1,...,T$}
        \IF{$b_t < b^*_{w}$}
            \STATE \textbf{action} $\gets$ \textit{Up}
        \ELSIF{$b_t =  b^*_{w}$}
            \STATE \textbf{action} $\gets$ \textit{Stay}
        \ELSE
            \STATE \textbf{action} $\gets$ \textit{Down}
        \ENDIF
        \ENDFOR
    \end{algorithmic}
\end{algorithm}





\section{System Overview}
\label{System Overview}

Rather than training \gls{rl} algorithms on the real database environment by collecting metrics in real-time, we choose to first collect training data that covers all possible states and then use it for model training. Firstly, this approach is more efficient in terms of computational time and used resources. Indeed, after changing the buffer size, a waiting time is necessary for system stabilization before evaluating the state and the latency. Moreover, by covering all the possible states once, the cost of the test of several algorithms is significantly reduced.
Secondly, this approach allows to evaluate the Oracle for calculating the reward of trained models.
That is why our \gls{rl} based database tuning system uses three phases: Data Collection, Exploration (model training), and Exploitation (real-time tuning).


\subsection{Data Collection Phase}
\label{data_collection_phase}

The data collection phase is designed to gather data for the subsequent model training and testing (Algorithm \ref {alg:DataCollection}). 
The data consists of a series sextuplets \( \langle w, b_t, s_t, l_t,l^*,b_w^* \rangle_{w \in \mathcal {W},t \in \{1,...,T\}} \), where:

\begin{itemize}
    \item \( w \in \mathcal {W}\) is the workload,
    \item \( b_t \in [B_{min},B_{max}]\) denotes the buffer size,
    \item \( s_t \in \mathcal {S}\) is the database state, which is reflects the observed workload $w_t$,
    \item \( l_t > 0\) is the latency,
    \item $l^*$ the target latency,
    \item $b^*_{w}$ is the optimal buffer size for workload $w$.
\end{itemize}

After the data generation, the data store is split into three datasets: one for training the models, one for validating the parameters, and one for testing.



\begin{algorithm}
	\caption{Data Collection}
	\label{alg:DataCollection}
	\begin{algorithmic}[1]
    \FOR {$w \in \mathcal{W}$}
        \STATE $t \leftarrow 0$
		\FOR {$b_t\leftarrow B_{max}$ \textbf{to}  $B_{min}$} 
            \STATE Set buffer size to $b_t$
            \STATE Execute $w$
            \STATE Measure $s_t,l_t$
            \STATE $t \leftarrow t+1$
        \ENDFOR
        \STATE $b^*_w \leftarrow \min_{i \in \{0,1,...,t\}, l_{i} \leq l^*} b_i$
         \STATE $l^*_w \leftarrow \max_{i \in \{0,1,...,t\}, l_{i} \leq l^*} l_i$
        \FOR {$i \in \{0,1,...,t\}$}
            \STATE Write $w,b_i,s_i,l_i,l^*, b^*_{w}$ in the data store
        \ENDFOR
	\ENDFOR
    \STATE Split data store according to $w$ in training, validation and test datasets.
	\end{algorithmic}
\end{algorithm}

\subsection{Exploration Phase}
\label{exploration}
The goal of the exploration phase is to optimize the policy parameters by training the selected algorithm  and rigorously evaluating the resulting policy’s performance.
In the exploration phase, a virtual DBMS tuning environment is built by replaying the datasets of sextuplets collected with Algorithm~\ref{alg:DataCollection}:
$ \langle w,\,b_t,\,s_t,\,l_t,\,l^*,\,b^*_w\rangle$. 
At each step, the agent’s action sets the virtual DBMS buffer size $b_t$. We then query the collected dataset with the full context tuple $\langle w,b_t,l^*,b^*_w\rangle$ to retrieve the corresponding state–metric pair $\langle s_t,l_t\rangle$.
This “memorize-and-replay” approach accelerates training and guarantees full reproducibility.



\begin{algorithm}
\caption{RL Exploration Phase}
\label{alg:rl_exploration}
\begin{algorithmic}[1]
  \STATE Initialize policy $\pi_0$ randomly
  \FOR{$w$ in $\mathcal{W}_{train}$}
    \STATE Set initial buffer size $b_0 \sim \mathcal{N_T}(\mu,\sigma,a,b)$
    \FOR {$t \leftarrow 1,...,T$}
      \STATE Observe $s_t$ 
      \STATE Select action $a_t \in \{\text{Down}, \text{\textit{Stay}}, \text{\textit{Up}}\}$ using $\pi_t$
      \STATE Update $b_t$ by applying $a_t$ in the environment
      \STATE The environment changes from the state $s_t$ to $s^{\prime}_t$
      \STATE Observe $s^{\prime}_t,l_{t}$ and compute reward $r_t$ according to equation \ref {eq:reward}
      \STATE Update $\pi_t$ using $(s_{t}, a_{t}, r_{t},s^{\prime}_t)$

    \ENDFOR
  \ENDFOR
  \STATE Evaluate and Store policy $\pi_{\mathcal {W},l^*}$ 
\end{algorithmic}
\end{algorithm}

Algorithm~\ref{alg:rl_exploration} outlines the exploration process. 
To train the agent to interact frequently with values near \(b^*_{w} \), the starting point is initialized by selecting a buffer size from a truncated Gaussian distribution $\mathcal{N_T}(\mu,\sigma,a,b)$, where parameters $\mu=b^*_{w}$ is the mean,$\sigma$ is the standard deviation $a=\frac{\,B_{\min} - b^*_{w}\,}{\sigma}$ is the lower bound, and $b=\frac{\,B_{\max} - b^*_{w}\,}{\sigma}$ is the upper bound.
\(T\) is chosen to be sufficiently large so that the agent can optimize the buffer for each workload $w$. 



\subsection{Exploitation Phase}

After the policy has been trained and evaluated, the policy is stored in a model warehouse. 
The point of a model warehouse is to store policies trained under different scenarios and different levels of \gls{sla}, and hence, which have different behaviors. The model warehouse provides different solutions for different needs.
In the real-time exploitation phase, when the DBMS administrator makes a tuning request, he has to choose a model corresponding to its DBMS configuration in the model warehouse. The controller loads the selected model and starts to collect the metrics (DBMS state) from the DBMS, which is typically under the production workload from an application.
The \gls{rl} algorithm tunes the buffer size based on the observed state by applying modifications to the production DBMS. This process is repeated to optimize RAM utilization while preventing SLA violations.

\section{Experimentation}
\label{exps}

In this section, the performance and the efficiency of \emph{MicroTune} are evaluated under different core algorithms and baselines. We show and compare their effectiveness by evaluating their capability to reduce RAM utilization while respecting \gls{sla} constraints compared to the baselines. To evaluate the performances of different \gls{rl} algorithms, and to tune the hyperparameters of these algorithms (Section \ref{Hyperparameter Tuning}), the following metrics are used:


\paragraph{1. Total number of \gls{sla} Violations.}
Let \(V_{SLA}\) denotes the total number of SLA violations observed.
This metric evaluates the number of times where $l_t$ under $b_t$ adjusted by the policy fails to meet the \gls{sla} requirement $l^*$. It serves as an indicator of the agent's capability to satisfy \gls{sla} constraints. The $V_{SLA}$ is defined as follows:
\begin {equation}
V_{SLA_T}(\pi)={\sum_{t=1}^T \mathds{1}(l_t > l^*)}.
\end {equation}


\paragraph{2. Cumulative RAM Utilization.}  
The cumulative RAM Utilization metric, denoted $C_{RAM}$, calculates the total buffer size utilized across all time steps. It is used to assess the agent's ability to minimize buffer size while managing workload demands effectively. This metric is defined as follows:
\begin {equation}
C_{RAM_T}(\pi)=  {\sum_{t=1}^T b_t}.
\end {equation}

\paragraph{3. Normalized Distance to optimal policy $d_T(\pi,\pi^*)$.}

The main metric we use to evaluate the overall effectiveness of a given policy as it leverages both the $V_{SLA}$ and the $C_{RAM}$ compared to the optimal policy (Algorithm \ref {alg:Oracle}).
The normalized distance of a given policy $\pi$ to the optimal policy $\pi^*$ on the dataset of length $T$ is defined as following:

\begin{equation}
d_T(\pi,\pi^*)=
\frac{\sum_{t=1}^T \left(\mathds{1}(l_t > l^*) - \mathds{1}(l^*_{t} > l^*)\right)}{\sum_{t=1}^T \mathds{1}(l^*_{t} > l^*)}
+ 
\frac{\sum_{t=1}^T \left(b_t - b^*_{t}\right)}{\sum_{t=1}^T b^*_{t}},
\label{normalized_distance}
\end{equation}
where $b^*_t$ and $l^*_t$ are respectively the buffer size and the latency chosen by the optimal policy (Algorithm \ref{alg:Oracle}), and $l^*$ is the target latency.
This normalized distance serves as a general metric for evaluating and comparing policies by measuring how closely each one approximates the optimal policy.

\subsection{Workloads and Datasets}
\label{workoads}

We used MariaDB 11.1.3 and the Sysbench benchmark to test and evaluate our work. MariaDB is deployed on Flexible Engine~\cite{orangeFlexibleEngine} using machines with 4 vCPU cores and 16 GB of RAM (model c6.xlarge.4). We collected states and performances of a total of 592 different workloads generated by Sysbench executed on MariaDB. It takes approximately 37.5 minutes to collect the data points for one workload. The setup relies on 3 machines, one for the benchmarking tool as a database client, one for the DBMS server, and one to pilot operations and collect data. The whole run continued for 16 days to collect data for all 592 workloads.
A workload $w$ within our database tuning context is characterized by the following factors: \begin{itemize} \item The database size evolves from 128MB up to 8GB with a number of tables ({5, 15, 22, 30, 40, 50}) and number of rows per table ({1000000, 300000, 100000}). \item The number of simultaneous clients connected to the database ({1, 2, 3, 4, 5, 6, 7, 8, 9, 10, 11, 12}). \item The distribution of data access, with four types currently available: special, Gaussian, uniform, and Pareto. The uniform distribution allows for more extensive and equal access to all data during a load, while Pareto focuses on accessing a smaller subset of the dataset more intensively. \end{itemize}

 We collect 95 percentile latency for every second from the benchmark tool Sysbench. 
Due to system fluctuations and network issues, we may observe outliers of the measured latency in our datasets even with 95\% percentile. These outliers may significantly skew the training process and lead to inaccurate learned policies of the agent. To address this issue, a simple moving average smoothing is applied:
\begin {equation}
l'_i = \frac{1}{3}(l_{i-1} + l_i + l_{i+1}).
\end {equation}

In Algorithm \ref{alg:DataCollection}, we initialize the Sysbench benchmark with the DBMS set to a buffer size of 8G. Then,
we decrease the DBMS buffer size after the collection of each tuple \( \langle w, b_t, s_t, l_t,l^*,b^*_{w} \rangle \), with a granularity of 128MB and repeat this process, until it reaches the minimum value, which is 128MB. Then, we change the workload and repeat the process. During this process, a total of 37888 tuples \( \langle w, b_t, s_t, l_t,l^*_{w},b^*_{w} \rangle \) are generated. We set the target latency \(l^*\) to 20\,ms, a DBA-recommended compromise between memory usage and performance. $b^*_w$ is evaluated on each workload using a conservative tuning of the parameter in equation (\ref {eq:OIA}): $\delta=0$.

The collected dataset is partitioned according to the workloads into three disjoint subsets: 60\% for training (used in the exploration‐phase model fitting), 20\% for validation (for RL hyperparameter tuning), and 20\% for testing. Each subset is then used to instantiate its own virtual DBMS tuning environment (see Section~\ref{exploration}). 
In Algorithm \ref {alg:rl_exploration}, the number of steps for each workload is set to $T=68$.

\subsection{Tested algorithms }

In this section, we present the algorithms evaluated in our study. To ensure comprehensive coverage, the evaluation includes several deep \gls{rl} algorithms with Stable-Baselines3~\cite{SB3} (SB3) implementation,
alongside contextual bandit algorithms.
Additionally, Horizontal Pod Auto-scaling (HPA)~\cite{kubernetes_hpa} is assessed as a baseline, and some other rule-based baselines are also proposed.

Rule-based baselines do not require a training phase but always rely on their specific inputs: for example, the \emph{Basic} and \emph{HPA} baselines require \(l_t\), the \emph{Miss Ratio} baseline depends on the miss ratio distribution, and the Optimal Policy is contingent on \(b^*_w\). RL baselines require a training phase with their specific inputs, but once trained, in the exploitation phase, they can be used without the latency $l_t$ as reward computation is not needed anymore.  Baselines that do not rely on the latency in the exploitation phase are more reflective of most industrial scenarios, where collecting real-time latency data is either impractical or prohibitively expensive.

\subsubsection{Baseline: Optimal Policy.}
\label{oracle}
The algorithm that makes perfect actions as described in Algorithm \ref{alg:Oracle}. This baseline cannot be used in real time in the exploitation phase.


\subsubsection{Baseline: Basic.} A straightforward method to manage buffer sizes is to mimic the intuitive actions a human operator might take in real time. 
Given a workload, one could compare at each interval, $l_t$ against $l^*$ with a tolerance $\epsilon$ and adjust the buffer size accordingly. In our experimentation, the tolerance is set to $\epsilon = 0.1$.


\subsubsection{Baseline: Strategy HPA.}
HPA~\cite{kubernetes_hpa} is an algorithm used in Kubernetes and is responsible for managing the number of Pods for an application facing a fluctuating workload. HPA determines the number of Pods based on the ratio between the desired and the current metric values, as follows:
\begin {equation}
\label{eq:hpa}
\text{Optimal Replicas} = \lceil \text{current Replicas} \times \left( \frac{\text{current Metric Value}}{\text{desired Metric Value}} \right) \rceil.
\end {equation}

Inspired by HPA, in the context of memory allocation for \emph{MicroTune}, the equation \ref{eq:hpa} has been adapted to compute, at each time step $t$, the optimal buffer size denoted $b^*_{HPA_t}$, as 
\begin {equation}
\label{eq:hpa_updated}
b^*_{HPA_t} =  \lceil b_t \times \left( \frac{l_t}{l^*} \right) \rceil. \
\end {equation}

If $b_t$ is less than $b^*_{HPA_t}$, then the action \textit{Up} is selected, if $b_t$ is within the range of \([b^*_{HPA_t}, b^*_{HPA_t} + 128\,\text{MB}]\), then the action \textit{Stay} is chosen, else the action \textit{Down} is applied. This strategy allows us to adjust $b_t$ step by step to $b^*_{HPA_t}$ while updating the $b^*_{HPA_t}$ for every time steps $t$. The range of \([b^*_{HPA_t}, b^*_{HPA_t} + 128\,\text{MB}]\) allows the HPA baseline to over-allocate a little RAM space to ensure that \gls{sla} is met. 

As shown in equation \ref{eq:hpa_updated}, it is important to note that this algorithm assumes a strong linear relationship between the latency and the buffer size. The collected data suggests that this assumption may hold in certain cases, but is limited.

\subsubsection{Baseline: Miss Ratio.}
Similar to \textsc{iBTune}~\cite{ibtune}, we implement a rule-based tuning baseline using the miss ratio, defined as the fraction of read requests not served from the buffer pool and thus requiring disk access. In InnoDB, this is computed as \texttt{innodb\_buffer\_pool\_reads} over \texttt{innodb\_buffer\_read\_requests}, observed in state metrics $s_t$. A lower miss ratio implies effective caching, while a higher ratio indicates insufficient buffer size. Unlike \textsc{iBTune}, which directly targets a desired miss ratio, we map it empirically to a latency threshold that satisfies the \gls{sla}. Since this mapping depends on workload and hardware, the method is only a comparison baseline. 



\subsubsection{RL agorithms.}
We compare against several off-the-shelf algorithms: LinUCB~\cite{LinUCB}, a contextual bandit with upper confidence bounds applied in cloud scaling~\cite{david} and used here with the disjoint-arm variant~\cite{KFOOFW}; 
Proximal Policy Optimization (PPO)~\cite{PPO}, which stabilizes policy gradient updates via a clipped surrogate objective; Deep Deterministic Policy Gradient (DDPG)~\cite{ddpg}, an actor-critic algorithm also used in {\sc CDBtune}~\cite{CDBtune}, Deep Q-Networks (DQN)~\cite{dqn}, which approximate $Q(s,a)$ with deep nets using replay buffers and target networks, and Advantage Actor-Critic (A2C)~\cite{a2c}, a synchronous actor-critic approach leveraging the advantage function for stable updates.

\subsection{Reward Shaping}

The reward function is designed to encourage the agent to increase the buffer size
when actual performance does not meet SLA requirements; encourage decreasing the buffer size when SLA requirements are already met; reward the
agent for maintaining a certain buffer size that meets SLA requirements.

Let $d_+$ and $d_-$ be two sigmoid functions $\mathds{R^+} \rightarrow [0,1]$ (See supplementary materials for details), the reward function for each actions is defined as:

\begin{equation}\label{eq:reward}
r_t =
\begin{cases}
\bigl( 1-\alpha) d_+(l_{t}),\, \beta (1-d_+(l_{t})),\, -\alpha d_+(l_{t}) \bigr) & \text{if } l_t < l^*, \\[6pt]
\bigl( 1-\alpha)d_-(l_{t}),\, \beta d_-(l_{t}),\, -\alpha d_-(l_{t}) \bigr) & \text{if } l_t \geq l^*,
\end{cases}
\end{equation}

where weight parameters $\alpha,\beta \in [0,1]^2$ allows balance between minimizing SLA violations and optimizing RAM utilization.

\begin{figure*}[htp!]
    \centering
    \includegraphics[width=0.49\linewidth]{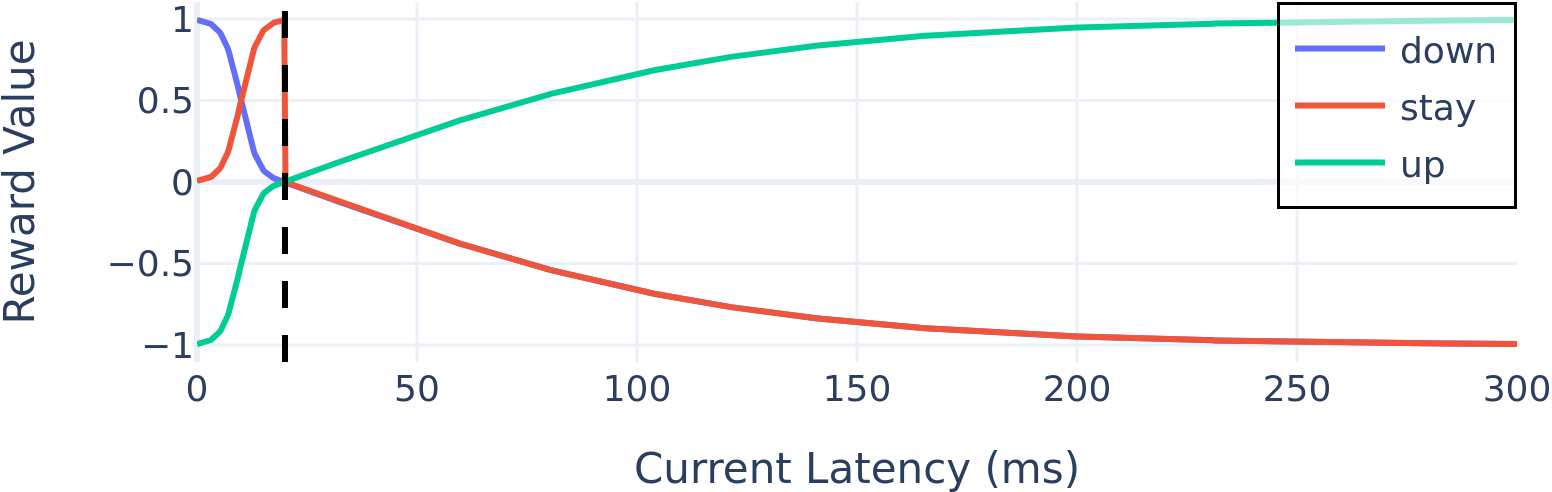}
    \includegraphics[width=0.49\linewidth]{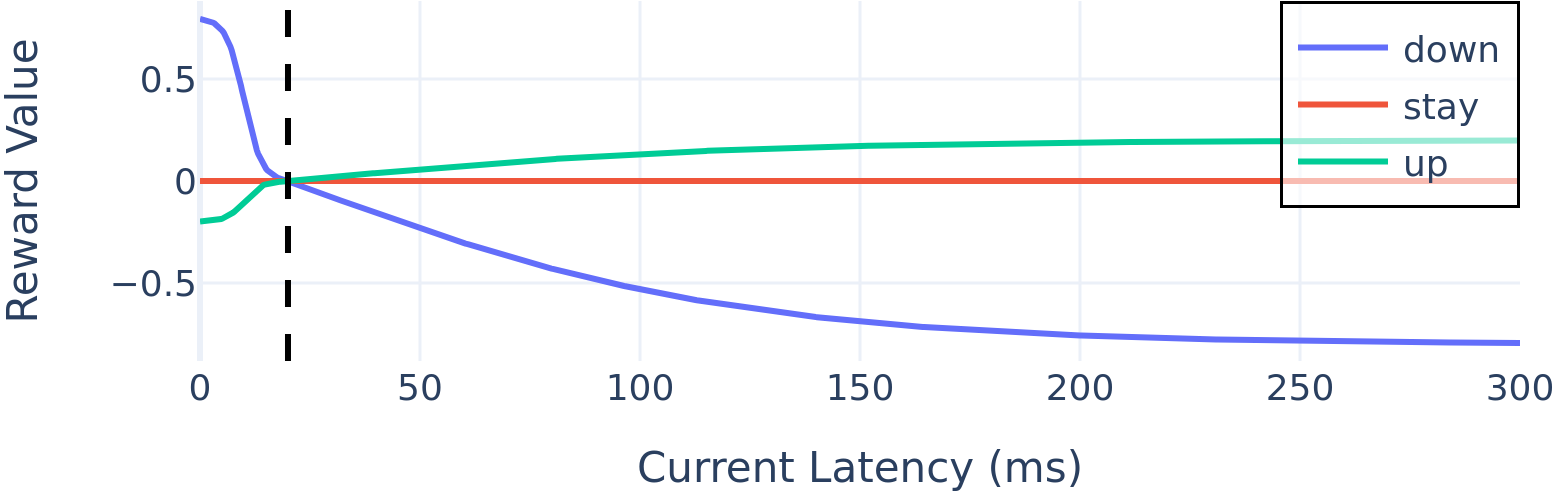}
    \caption{Reward evolution. On left without applying weights to the rewards ($\alpha=0$,$\beta=1$). On right using the best identified parameters value $\alpha=0.2$,$\beta=0$.}
    \label{fig:reward_function}
\end{figure*}

Figure \ref{fig:reward_function}, on left, depicts the evolution of the rewards for each actions depending on the observed latency with  $l^*=20ms$. Without applying weight parameters, one can note that the rewards evolutions are not well separated between the actions \textit{Down} and \textit{Stay} when $l_t > l^*$ leading to an uncertainty.
We performed a grid search over the coefficients \(\alpha\) and \(\beta\) which identified \(\alpha = 0.2\) and \(\beta = 0\) as the optimal values.
that identified \(\alpha = 0.2\) and \(\beta = 0\) as the optimal values.
Figure \ref{fig:reward_function}, on right, illustrates the reward evolution with the optimal weight parameters applied. We can observe that reward evolutions are well separated, the '\textit{Up}' or '\textit{Down}' action produces a relatively larger reward than the \textit{Stay} action and, when $l_t = l^*$, the agent receives a better reward for choosing \textit{Stay} action rather than \textit{Up}, which would decrease latency, or \textit{Down} action, which would increase it and therefore lead in both cases to a negative reward. 

\subsection{ \gls{rl} Hyperparameter Tuning}
\label{Hyperparameter Tuning}
RL performance depends critically on hyperparameters such as learning rate (which controls the step size during parameter updates), batch size (the number of samples processed in one iteration), and discount factor gamma (the discount factor that balances immediate and future rewards)—to avoid unstable training or poor convergence~\cite{tuning}. We used Optuna~\cite{Optuna} with its Tree-structured Parzen Estimator (TPE) \cite{TPE}, which models separate distributions for promising and poor values and selects new samples by maximizing expected improvement.  To ensure a diverse and well-informed search, the optimization process is initialized with random sampling for the first few trials before applying TPE.
We use the Normalized Distance to optimal policy (see Equation~\ref{normalized_distance}) as the performance metric.  



\begin{table*}[htp!]
\centering
\caption{Performance of algorithms on test set}
\begin{tabular}{|l|ll|ll|ll|}
\hline
                   & \multicolumn{2}{l|}{SLA Violations}                                       & \multicolumn{2}{l|}{\makecell{Cumulative RAM\\Utilization ($\text{TB}$)}}                            & \multicolumn{2}{l|}{\makecell{$d_T(\pi,\pi^*)$}}                               \\ \cline{2-7} 
\multirow{-2}{*}{} & \multicolumn{1}{l|}{Train}                       & Test                       & \multicolumn{1}{l|}{Train}                       & Test                         & \multicolumn{1}{l|}{Train}                            & Test                             \\ \hline
LinUCB             & \multicolumn{1}{l|}{2651 ± 140}              & 2196 ± 112           & \multicolumn{1}{l|}{86.6 ± 0.6}                  & 78.7 ± 0.7                 & \multicolumn{1}{l|}{0.6 ± 0.07}                        & 1.4 ± 0.1                        \\ \hline
DQN                & \multicolumn{1}{l|}{2055 ± 54}               & 1288 ± 122           & \multicolumn{1}{l|}{92.1 ± 1.6}                  & 82.9 ± 1.45                 & \multicolumn{1}{l|}{{\color[HTML]{CB0000} 0.38 ± 0.0}} & 0.6 ± 0.1                        \\ \hline
A2C                & \multicolumn{1}{l|}{2040 ± 24}               & 1335 ± 101           & \multicolumn{1}{l|}{93.5 ± 1.5}                  & 83.6 ± 1.55                 & \multicolumn{1}{l|}{{\color[HTML]{CB0000} 0.39 ± 0.0}} & {\color[HTML]{CB0000} 0.7 ± 0.1} \\ \hline
PPO                & \multicolumn{1}{l|}{1946 ± 22}               & 1250 ± 44            & \multicolumn{1}{l|}{94.8 ± 1.2}                  & 86.3 ± 1.05                 & \multicolumn{1}{l|}{{\color[HTML]{CB0000} 0.36 ± 0.0}} & {\color[HTML]{CB0000} 0.66 ± 0.1} \\ \hline
DDPG               & \multicolumn{1}{l|}{2276 ± 381}              & 1703 ± 471           & \multicolumn{1}{l|}{92.0 ± 3}                  & 82.5 ± 2.88                 & \multicolumn{1}{l|}{0.5 ± 0.16}                        & 1.0 ± 0.4                        \\ \hline
Basic              & \multicolumn{1}{l|}{4293}                        & 3312                       & \multicolumn{1}{l|}{75.4}                        & 66.4                        & \multicolumn{1}{l|}{1.26}                             & 2.38                             \\ \hline
HPA                & \multicolumn{1}{l|}{3808}                        & 3366                       & \multicolumn{1}{l|}{74.8}                        & 63.4                        & \multicolumn{1}{l|}{1.00}                             & 2.44                             \\ \hline
Miss Ratio         & \multicolumn{1}{l|}{5914}                        & 5907                       & \multicolumn{1}{l|}{91.7}                       & 80.8                        & \multicolumn{1}{l|}{2.33}                             & 5.24                             \\ \hline
Static          & \multicolumn{1}{l|}{}                        &                     & \multicolumn{1}{l|}{}                       &                        & \multicolumn{1}{l|}{}                             &                              \\ 
80\% RAM         &  \multicolumn{1}{l|}{1792}                      &     0               &    \multicolumn{1}{l|}{217}                    &      217                &                  \multicolumn{1}{l|}{1.87}            &           1.33                \\ 
50\% RAM         & \multicolumn{1}{l|}{2816}                        &          1024              & \multicolumn{1}{l|}{136}                       &     136                  & \multicolumn{1}{l|}{1.31}                             &     1.13                        \\  \hline

Optimal Policy             & \multicolumn{1}{l|}{{\color[HTML]{CB0000} 1914}} & {\color[HTML]{CB0000} 984} & \multicolumn{1}{l|}{{\color[HTML]{CB0000} 73.8}} & {\color[HTML]{CB0000} 65.12} & \multicolumn{1}{l|}{0}                                & 0                                \\ \hline
\end{tabular}

\label{table:result_table}
\end{table*}

\subsection{Experimental results on test set}
\label{effectiveness}

The evaluated algorithms range from deep learning methods to simple heuristics. All experiments were conducted under identical 
hardware configurations to ensure fair comparisons. 
Table~\ref{table:result_table} reports the performance of all algorithms under their best hyperparameter settings, each averaged over $10$ random seeds generating different partitions for training and evaluation datasets. 
Algorithms are evaluated on unseen test workloads, thus assessing both their ability to learn under training conditions and to generalize to workload shifts in realistic settings.
For each tested workload, one run is done with a starting buffer size fixed to its minimal value $B_{min}=128$MB, and one run is done with a starting buffer size fixed to its maximal value $B_{max}=8$GB.
Because test workloads include small initial buffer sizes, the tested problem is particularly difficult: even the optimal policy leads to temporary SLA violations when starting with a buffer size much smaller than the optimum.
The optimal policy establishes the lower bound on both SLA violations and memory cost. 
In addition to the adaptive baselines, we also compared the algorithms' performance against static buffer sizes fixed at 80\% (6,528\,MB) and 50\% (4,096\,MB) of the machine's total RAM, a strategy more commonly used in production environments.

For the static buffer sizes, we observe that fixing RAM at 80\% significantly reduces SLA 
violations; however, it also leads to substantial memory over-utilization compared to adaptive 
algorithms, resulting in a higher normalized distance than RL-based methods. Conversely, 
setting RAM at 50\% reintroduces SLA violations while still consuming considerable memory. 
These results underscore the difficulty of choosing an appropriate fixed buffer size in 
production to balance SLA compliance and resource efficiency, thereby motivating the need 
for adaptive approaches.
For adaptive solutions, RL methods clearly outperform rule-based baselines, even without direct information about latency. Among them, PPO, DQN, and A2C achieve performance closest to the optimal policy, while LinUCB lags behind. Their success stems from leveraging sequential, non-linear feedback, whereas DDPG struggles with our discrete bounded action space and sensitivity to hyperparameters. Rule-based baselines (Basic, HPA, Miss Ratio) suffer from oscillations, poor SLA compliance, or large deviations from the optimal policy, confirming the limits of non-learning approaches.




\section{Experiments in real-time}
\label{online}

\begin{figure*}[htp!]
    \centering
    \includegraphics[width=1\linewidth]{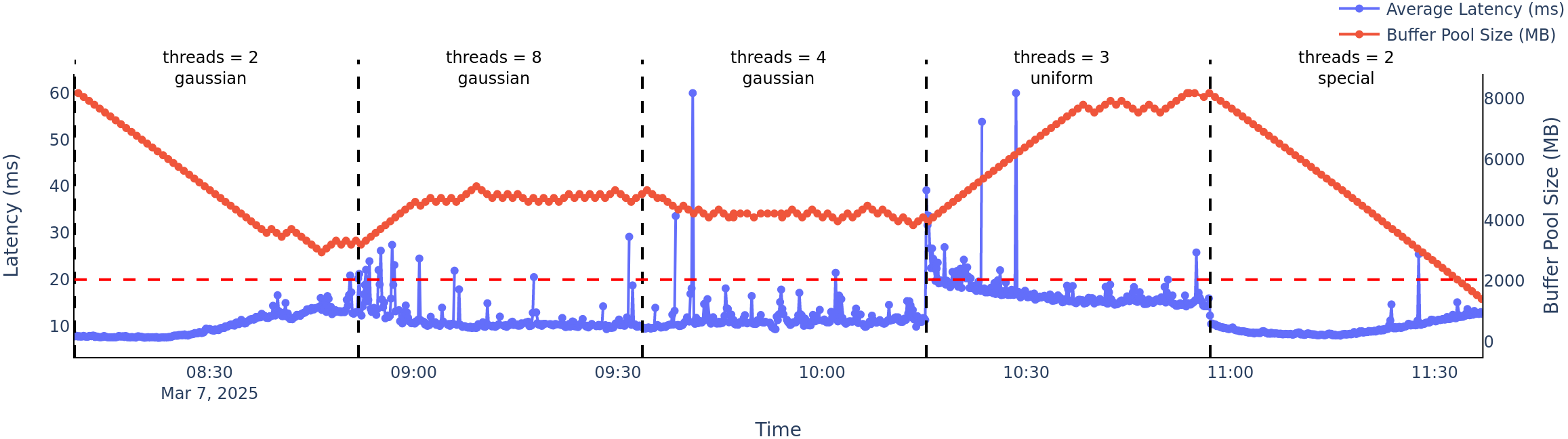}
    \caption{\emph{MicroTune} Online Tuning with varying workloads with A2C agent}
    \label{fig:online_tuning}
\end{figure*}

In this section, we demonstrate how the trained \emph{MicroTune} agent, using the A2C algorithm (one of the top-performing algorithms), tunes a \textbf{live database in real-time} to minimize $C_{RAM}$ and $V_{SLA}$. Compared with a simulated environment, real-time tuning introduces additional noise, making it more challenging for the agent to take appropriate actions to meet the latency target.

Figure~\ref{fig:online_tuning} illustrates the online tuning process for a Sysbench workload (50 tables × 1M rows) with a 20 ms latency goal. We change thread counts and access patterns periodically to mimic production environments. 
The red line is buffer‐pool size; the blue line is the 10-second moving average of latency; black dotted lines indicate workload shifts. We run this on the same machine used for training, and for demonstration set MicroTune’s buffer‐size update interval to 30 s, production deployments would use a lower frequency for stability.

\textbf{Phase 1: Low Workload (2 threads, Gaussian distribution)}
Initially, the workload runs with 2 threads under a Gaussian distribution, and the buffer pool is initialized at 8\,GB. The resulting latency is far below the 20\,ms target, indicating that the buffer pool is over-allocated. The agent recognizes this and gradually downsizes the buffer until it stabilizes around 4\,GB. Although some outliers briefly exceed 20\,ms during the transition, these short spikes are normal in a live system subject to background noise.

\textbf{Phase 2: Higher Workload (8 threads)}
Next, the number of threads is increased to 8 to stress the database further. The latency begins to exceed the target more frequently, prompting the agent to increase the buffer pool size. As the buffer grows, the latency stabilizes again below 20\,ms.

\textbf{Phase 3: Moderate Workload (4 threads)}
Then, the workload concurrency is reduced to 4 threads. The agent detects that the system can operate with a smaller buffer and reduces the size from approximately 4.8\,GB to 4\,GB without incurring SLA violations.

\textbf{Phase 4: Different Access Pattern (3 threads, Uniform distribution)}
The workload is then switched to 3 threads with a uniform distribution, which is more intensive than the Gaussian pattern. Latency immediately spikes above the threshold, and the agent responds by increasing the buffer pool to 8\,GB. This brings latency back under 20\,ms, demonstrating the agent’s ability to react to heavier workloads.

\textbf{Final Phase: Lighter Workload (2 threads, Special distribution)}
Finally, the workload is lowered to 2 threads with a “special” distribution, resulting in a relatively light load. The agent observes decreased demand and reduces the buffer size accordingly.

Overall, this live demonstration confirms the agent’s capability to handle dynamic changes in workload while keeping latency within acceptable bounds and reducing the buffer pool size when possible despite occasional short-lived spikes.




\section{Conclusion}
\label{conclusion}
In this paper, we introduced \emph{MicroTune}, an online tuner that continuously monitors database metrics and optimizes memory usage while respecting SLA constraints. 
We formally stated the problem of optimizing the buffer size under SLA constraints and an Oracle is deduced from it. 
\gls{rl} is used to reach the buffer size of the Oracle for each workload.
Rather than learning the policy online, which is expensive in time and may raise safety issues, we start with data collection for workloads with various characteristics, which enables us to evaluate the optimal buffer size for each workload.
Then, we defined an optimal policy (target policy) that can be computed offline on the data, which allows us to reach Oracle's performance step by step. The data allows us to train the \gls{rl} algorithms, and to perform hyperparameter tuning to select the best-trained policy. Then, the best-trained policy can be used in real-time to tune the buffer size of a DBMS under SLA constraints.
We defined some baselines, which could be used as an alternative to \emph{MicroTune}, and we tested the state-of-the-art reinforcement learning algorithms. The results coming from our experiments are clear: while \gls{rl} algorithms do not use latency, they dominate all baselines, even those that use it.
Finally, we selected the best-trained policy, based on A2C, and we successfully tested it in real time using realistic workloads that change over time.
In future works, We aim to deploy \emph{MicroTune} on production systems to log data for counterfactual policy learning~\cite{Zenati2023} and to further investigate how to keep the system stable while choosing tuning intervals that minimize performance fluctuations.

\bibliographystyle{splncs04}
\bibliography{sample-base}

\appendix

\section{Experiments replay}
The source code used in the present work is available at:

https://github.com/anonpublisher330/anon1

\section{Reward Function}
Using different functions for each action allow to favor different behaviors: for instance rewarding more the action \textit{Up} leads to more conservative policies minimizing SLA violations at the expense of minimizing the buffer size, while rewarding more action \textit{Down} leads to more risky policies in terms of SLA violations for minimizing the buffer size. 
Let $d_+$ and $d_-$ be functions $\mathds{R_+} \rightarrow [0,1]$:

\begin{align}
d_+(l_{t}) =
&= \frac{1}{1 + \exp\!\Biggl(-10\Bigl(\frac{l^*- l_t}{l^*} - 0.5\Bigr)\Biggr)}, \\[6pt]
d_-(l_{t}) 
&= 2\left(\frac{1}{1 + \exp\!\Bigl(\,\frac{l^*- l_t}{-50}\Bigr)} - 0.5\right).
\end{align}

\begin{figure}
    \centering
    \includegraphics[width=1\linewidth]{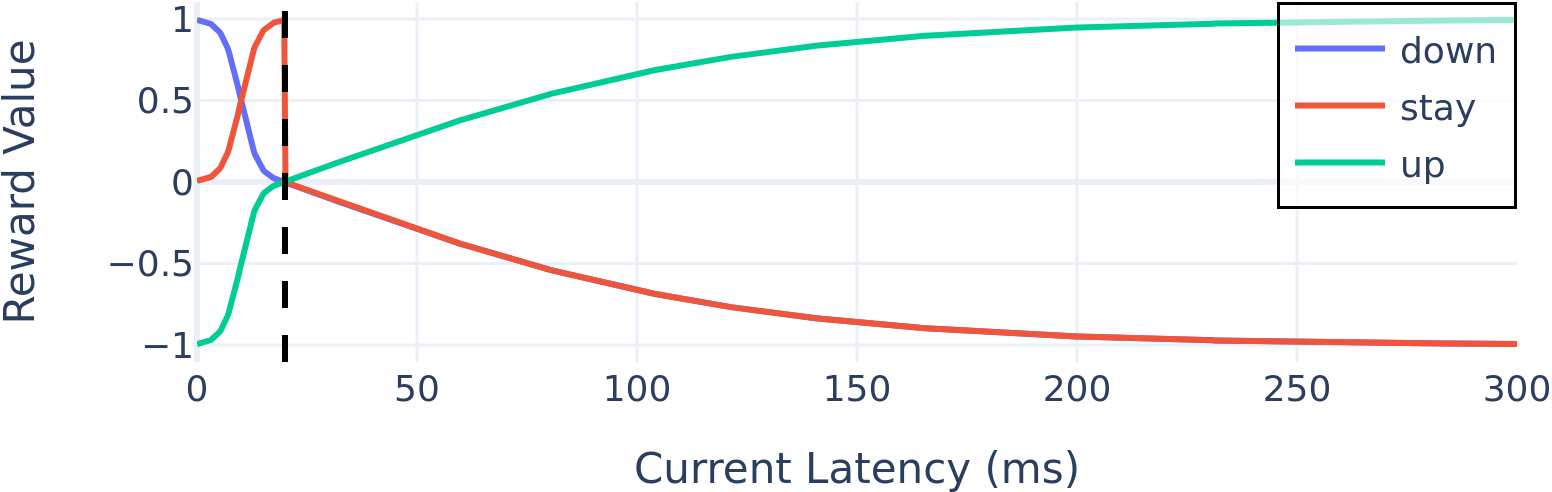}
    \caption{Shape of the reward function for each action}
    \label{fig:reward_function_original}
\end{figure}

The slopes of these scaled sigmoid functions are graphically selected (Figure \ref {fig:reward_function_original}). The reward for actions \((\text{\textit{Down}}, \text{\textit{Stay}}, \text{up})\) is defined as:

\begin{equation}\label{eq:reward}
r_t =
\begin{cases}
\bigl( 1-\alpha) d_+(l_{t}),\, \beta (1-d_+(l_{t})),\, -\alpha d_+(l_{t}) \bigr) & \text{if } l_t < l^*, \\[6pt]
\bigl( 1-\alpha)d_-(l_{t}),\, \beta d_-(l_{t}),\, -\alpha d_-(l_{t}) \bigr) & \text{if } l_t \geq l^*,
\end{cases}
\end{equation}

where $\alpha,\beta \in [0,1]^2$ are parameters that have to be optimized (see section \ref {Reward Engineering}).

The reward function with its associated slope parameters, is designed to encourage the agent to increase the buffer size when the observed performance fails to satisfy the \gls{sla} requirements. Conversely, increases in buffer size are penalized when the \gls{sla} requirements are already fulfilled. The agent is further rewarded for maintaining a buffer size that consistently satisfies the \gls{sla} constraints, while deviations leading to SLA violations incur substantial penalties. This reward structure encourages the system to self-correct and maintain performance within the optimal range.

\section{Reward Shaping }
Most \gls{rl} algorithms require extensive reward engineering to effectively solve challenging tasks~\cite{reward_shaping}. To refine the reward function, we constrain the parameters $\alpha$ and $\beta$ to lie within the interval [0, 1]. This calibration allows us to strike an appropriate balance between minimizing SLA violations and optimizing RAM utilization.

In the Equation \ref{eq:reward}, $\alpha$ governs the balance between the \textit{Up} and '\textit{Down}' actions. A smaller value of $\alpha$ implies a stronger penalty for executing a \textit{Down} action when SLA requirements are not met, while simultaneously rewarding a \textit{Down} action when they are satisfied. Conversely, a larger value of $\alpha$ favors rewarding an Up action in cases where SLA requirements are unmet and penalizes the Up action when they are met.
Meanwhile, the coefficient $\beta$ modulates the impact of the \textit{Stay} action relative to the other two actions. An elevated $\beta$ indicates that the agent gains more benefit from choosing to stay, whereas a lower $\beta$ suggests that the agent should place less emphasis on remaining in the same state.

\begin{figure}[htp!]
    \centering
    \includegraphics[width=.8\linewidth]{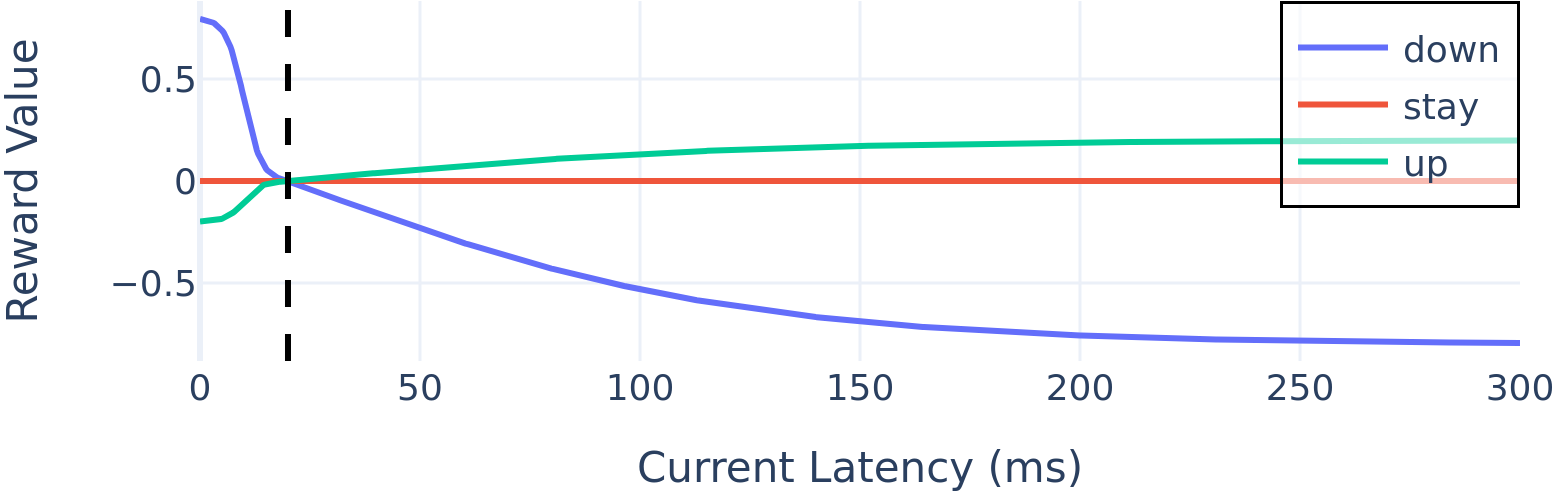}
    \caption{Visualization of Reward Function with coefficients}
    \label{fig:reward_function}
\end{figure}

\label{Reward Engineering}
\begin{figure}[htp!]
    \centering
    \includegraphics[width=.8\linewidth]{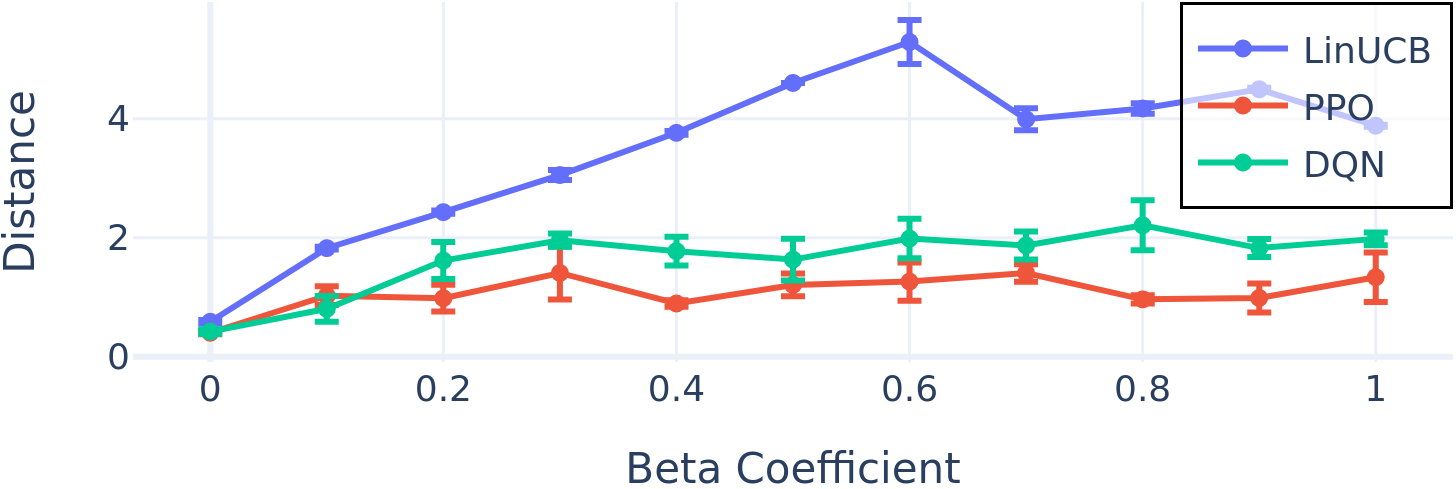}
    \caption{Mean and Std for Normalized Distance to the optimal policy under different Beta Coefficient}
    \label{fig:beta_sweep}
\end{figure}

We performed a grid search over the coefficients \(\alpha\) and \(\beta\), training on the 60\% training set and evaluating on the 20\% validation set. The objective was to minimize both the mean and standard deviation of the normalized distance to the optimal policy (see main paper Equation 4). The search identified \(\alpha = 0.2\) and \(\beta = 0\) as the optimal values. Consequently, the reward function reduces to:

\begin{equation*}
r_t =
\begin{cases}
\bigl( 0.8 \cdot d_+(l_{t}),\, 0,\, -0.2 \cdot d_+(l_{t}) \bigr) & \text{if } l_t < l^*, \\[6pt]
\bigl( 0.8 \cdot d_-(l_{t}),\, 0,\, -0.2 \cdot d_-(l_{t}) \bigr) & \text{if } l_t \geq l^*.
\end{cases}
\end{equation*}

Figure \ref{fig:reward_function} illustrates the reward function using the optimized values of $\alpha$ and $\beta$ when $l^*=20$ ms. The visualization reveals that, with the selected $\beta$, the \textit{Stay} action no longer yields the highest reward in every scenario; instead, either the 'Up' or '\textit{Down}' action produces a relatively larger reward. This suggests that the agent learns most effectively when the \textit{Stay} action has minimal impact, allowing it to remain in its current state without significant reward or penalty. Although consistently choosing \textit{Stay} would not maximize the immediate reward, it helps maintain system stability when the agent is uncertain about which action to take, ultimately leading to a lower long-term error.

To verify this finding, we conducted additional experiments by varying $\beta$ while keeping $\alpha$ fixed at its optimal value. As shown in Figure \ref{fig:beta_sweep}, the minimum distance across all three algorithms is achieved when $\beta = 0$.
This ensures the reward function strongly discourages growing the buffer through the larger weight on (\textit{Down}) while not excessively penalizing efforts to lower the buffer (\emph{up}). Simultaneously, it permits the agent to \textit{Stay} at the current state when no decisive action is warranted, ultimately yielding the best overall performance.

\end{document}